# A Heavy Higgs Particle in the TeV Mass Range ?

Chuan Liu [a*] with Karl Jansen[a] and Julius Kuti[a]

[a]Department of Physics 0319, University of California at San Diego, 9500 Gilman Drive, La Jolla, CA 92093-0319, USA

The first simulation results are presented on Higgs mass calculations in the spontaneously broken phase of the Higgs sector in the minimal Standard Model with higher derivative regulator. A heavy Higgs particle is found in the TeV mass range in the presence of a complex conjugate ghost pair at higher energies. The ghost pair evades easy experimental detection and a nonperturbative reinterpretation of the triviality Higgs mass bound becomes necessary.

## 1. Introduction

Following the systematic method that was developed earlier [1] we report our first simulation results on a heavy Higgs particle in the presence of a higher derivative regulator in the Higgs sector of the minimal Standard Model. The quantization procedure in Minkowski space-time with indefinite metric for the higher derivative Higgs Lagrangian,

$$\mathcal{L} = -\frac{1}{2}\Phi_\alpha \Box \Phi_\alpha - \frac{1}{2M^4}\Phi_\alpha \Box^3 \Phi_\alpha$$
$$- \frac{1}{2}m_0^2 \Phi_\alpha \Phi_\alpha - \lambda_0 (\Phi_\alpha \Phi_\alpha)^2 \; , \quad (1)$$

with $O(4)$ symmetry leads to the euclidean partition function [1]

$$\mathcal{Z} = \int \mathcal{D}\Phi \, \exp\left(-\int d^4x \, \mathcal{L}_E\right) \quad (2)$$

where $\mathcal{L}_E$ is obtained from Eq. (1) by replacing the Minkowski operator $\Box$ with the euclidean Laplacian.

The general particle content of the model is exhibited in the symmetric phase by the Hamiltonian in terms of creation and annihilation operators, $H = H_0 + H_{\text{int}}$,

$$H_0 = \sum \hbar \, [ \, \omega_{\mathbf{p}} \, a^{(+)}_{\alpha,\mathbf{p}} a^{(-)}_{\alpha,\mathbf{p}} + \Omega_{\mathbf{p}} \, b^{(+)}_{\alpha,\mathbf{p}} b^{(-)}_{\alpha,\mathbf{p}}$$
$$+ \Omega^*_{\mathbf{p}} \, c^{(+)}_{\alpha,\mathbf{p}} c^{(-)}_{\alpha,\mathbf{p}} \, ] \quad (3)$$

where the summation is over internal $O(4)$ indices and momentum modes. The dispersion of the original $O(4)$ particle is given by $\omega_{\mathbf{p}} = \sqrt{\mathbf{p}^2 + m_0^2}$ whereas the complex energy of the ghost state is $\Omega_{\mathbf{p}} = \sqrt{\mathbf{p}^2 + \mathcal{M}^2}$, with $\mathcal{M} = M e^{i\Theta}$; the complex phase $\Theta$ is a function of $m_0$ and $M$. The operator $a^{(+)}_{\alpha,\mathbf{p}}$ creates the physical particle with momentum $\mathbf{p}$ and energy $\hbar \omega_{\mathbf{p}}$. The operator $b^{(+)}_{\alpha,\mathbf{p}}$ creates a ghost state with momentum $\mathbf{p}$ and complex energy $\hbar \Omega_{\mathbf{p}}$; $c^{(+)}_{\alpha,\mathbf{p}}$ creates the complex conjugate ghost state. The interaction part of the Hamiltonian is proportional to the coupling constant and expressed in terms of creation and annihilation operators.

---

[*]Speaker at the Lattice 93 Conference, UCSD/PTH 93-39

In the broken phase the spectrum is more complex. There is a Higgs particle with mass $m_H$, and three massless Goldstone excitations with residual $O(3)$ symmetry. In addition, there is a complex conjugate pair of heavy ghost states with the internal quantum numbers of the Higgs particle, and a three-component complex ghost pair which transforms as $O(3)$ triplet according to the quantum numbers of the Goldstone particles. The heavy masses of the Higgs ghost and the Goldstone ghost are split by the symmetry breaking mechanism.

## 2. Lattice Action and Phase Diagram

For non-perturbative computer simulations of the higher derivative Lagrangian we introduce a hypercubic lattice structure. The lattice spacing $a$ defines a new short distance scale with the associated lattice momentum cut-off at $\Lambda = \pi/a$. We will have to work towards the large $\Lambda/M$ limit in order to eliminate finite lattice effects from the already regulated and finite theory. After rescaling the continuum field components by $\Phi_\alpha = \sqrt{2\kappa}\phi_\alpha$ the euclidean lattice action has the form

$$S_E = \sum_x [-\kappa\,\phi_\alpha(\Box + \frac{\Box^3}{M^4})\phi_\alpha + (1-8\kappa)\phi_\alpha\phi_\alpha$$
$$+\lambda\,(\phi_\alpha\phi_\alpha - 1)^2] \qquad (4)$$

where $\Box$ is the lattice Laplace operator, with the lattice spacing $a$ set to one for convenience. The mass parameter and coupling constant are related to the hopping parameter $\kappa$ and lattice coupling $\lambda$ by $m_0^2 = (1-8\kappa-2\lambda)/\kappa$, $\lambda_0 = \lambda/(4\kappa^2)$.

Stochastic algorithms exhibit severe critical slowing down for small values of the regulator mass $M$ in the range $M < 1$. In this regime the $M^{-4}\Box^3$ term leads to a dramatic broadening of the frequency spectrum of the Fourier modes. We developed a Hybrid Monte Carlo algorithm with fast Fourier acceleration which solved the problem of critical slowing down very effectively.

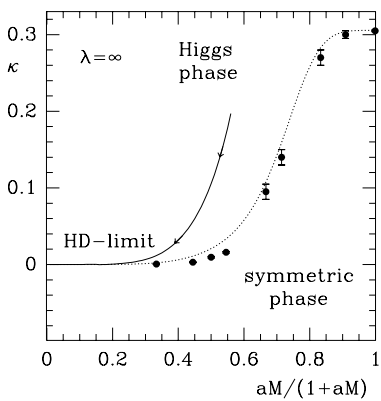

Figure 1. The phase diagram of the lattice model at infinite bare coupling. The dotted line is calculated in the large-N expansion. The solid line displays the fixed $M_R/m_H$ ratio towards the continuum limit of the higher derivative theory.

The lattice model exhibits two phases, as expected. The symmetric phase with full $O(4)$ symmetry is separated by a second order phase transition line from the broken phase which has a residual $O(3)$ Goldstone symmetry for every fixed value of $\lambda$ in the $(\kappa, M)$ plane. The phase diagram is shown in Fig. 1 in the $\lambda = \infty$ limit. Tuning the hopping parameter $\kappa$ to the critical line for fixed $M$ corresponds to the limit of trivial scalar field theory in the continuum. In this limit, the dimension eight operator $M^{-4}\phi_\alpha\Box^3\phi_\alpha$ becomes irrelevant and while $m_H \to 0$ in lattice units, $M_R/m_H \to \infty$, so that ghost effects from the higher derivative term disappear. In our notation $m_H$ is the physical Higgs resonance mass and $M_R$ designates the real mass parameter of the exact ghost pole location in the complex plane. The continuum limit of the higher derivative theory without the underlying lattice structure is equivalent to the tuning of $\kappa$ towards zero along a line of fixed $M_R/m_H$ ratio. In this limit, which is depicted in Fig. 1 as the solid line, the higher derivative term $M^{-4}\phi_\alpha\Box^3\phi_\alpha$ cannot be treated as an irrelevant operator in the Lagrangian.

## 3. Results and Discussion

Most of our computer simulations were performed on lattices which have cylinder geometry in the size range $16^3 \times 40$ to $20^3 \times 40$. On each lattice with a given parameter set 20,000 to 60,000 Hybrid Monte Carlo trajectories were accumulated in the Fourier transformed variables. On each time slice $\tau$ the radial Higgs field $\rho(\tau)$ and the $O(4)$ unit vector $n_\alpha(\tau)$ at zero three-momentum were defined as

$$\overline{\phi_\alpha}(\tau) \equiv \frac{1}{L^3}\sum_\mathbf{x} \phi_a(\tau,\mathbf{x}) \equiv \rho(\tau)n_\alpha(\tau)\ . \qquad (5)$$

The renormalized vacuum expectation value $v$ and the Higgs mass $m_H$ were determined from the correlation functions $\langle n_\alpha(0)n_\alpha(\tau)\rangle$ and $\langle \rho(0)\rho(\tau)\rangle$.

The fitting procedure requires a theoretical understanding of the energy spectrum which is organized in the large volume limit according to inverse powers of the linear spatial size $L$. The tower of $O(4)$ rotator states forms the most densely spaced energy levels with level spacing of the order $(v^2 L^3)^{-1}$. The rotator states dominate the correlation function $\langle n_\alpha(0)n_\alpha(\tau)\rangle$ [7]. The Goldstone excitations have a level spacing which is of the order $L^{-1}$. In the correlation functions at zero three-momentum only Goldstone pairs can contribute which have rather large energies in the volumes we are considering. The largest level spacings correspond to radial Higgs excitations with a finite energy gap $m_H$ above the ground state.

A simultaneous fit to $m_H$ and $v$ was performed on the data. The results are shown in Table 1 where all points correspond to a cylinder size of

Table 1
Simulation results at $M = 0.8$

| $\kappa$ | $\lambda_0/(10+\lambda_0)$ | $m_H$ | $v$ | $m_H/v$ |
|---|---|---|---|---|
| 0.056 | 0.761 | 0.40(2) | 0.058(2) | 6.8(4) |
| 0.056 | 0.761 | 0.32(2) | 0.045(1) | 7.1(6) |
| 0.105 | 0.185 | 0.30(2) | 0.066(2) | 4.6(4) |
| 0.115 | 0.086 | 0.21(3) | 0.084(2) | 2.5(4) |

$16^3 \times 40$ except for the second entry with size $20^3 \times 40$. The radial Higgs excitation in our simulations corresponds to a stationary state with real energy. Avoided level crossing with the lowest Goldstone pair of zero total momentum would only occur in larger spatial volumes. In our preliminary analysis the finite volume profile of the Higgs resonance was not developed. This rather crude approximation, where we identified the measured energy level of the radial excitation with $m_H$, is being subjected to systematic finite volume resonance analysis.

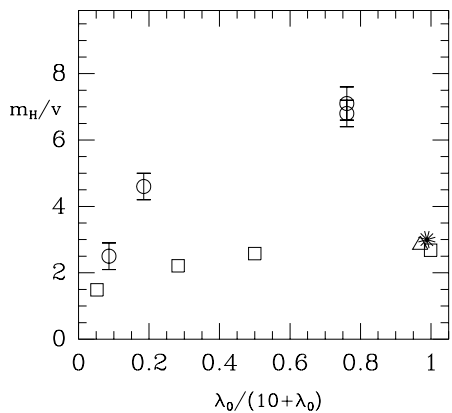

Figure 2. The circles are from our simulation results. They are compared with the simple $O(4)$ model on a hypercubic lattice [2] (squares), with Symanzik improved action on a hypercubic lattice [5] (star), and with dimension six interaction terms added on $F_4$ lattices [6] (triangle).

To compare our new results with earlier Higgs mass values, the ratio $m_H/v$ is plotted as a function of the bare coupling constant in Fig. 2. If our preliminary results will not change after a more developed analysis, $m_H \geq 1$ TeV with a ghost location in the multi-TeV region implies the existence of a strongly interacting Higgs sector, a scenario which was excluded in previous lattice studies. Our heavy Higgs mass values in the TeV range do not represent conventional triviality bounds [2–6] at fixed bare coupling constants. Since the ghost pair evades easy experimental detection without violating unitarity, Lorentz invariance, or any other sacred principles, a nonperturbative reinterpretation of the triviality Higgs mass bound becomes necessary.

To test the strongly interacting Higgs sector, we also determined the renormalized coupling constant in the symmetric phase. The preliminary results show stronger couplings in comparison with the lattice regulated $O(4)$ model.

4. Acknowledgements

This work was supported by the DOE under Grant DE-FG03-90ER40546 and by the Texas National Research Laboratory Commission under Grant RGFY93-206. The simulations were done at Livermore National Laboratory with DOE support for supercomputer resources.